\documentclass[11pt,a4paper]{article}
\usepackage{graphicx,color,times}

\usepackage{amsfonts}
\usepackage{amsmath,amssymb}
\usepackage{cite}

\usepackage{amsthm}

\makeatletter

\@addtoreset{equation}{section}
\makeatother
\let\refOld\ref
\renewcommand{\ref}[1]{(\refOld{#1})}

\newcommand{\tr}{{\rm tr~}}

\newcommand{\superp}[2]{\genfrac{}{}{0pt}{}{#1}{#2}}

 \def\d{\delta}
 
 \def\Re{{\rm Re ~}}
 
 \def\p{\partial}
 
 \def\a{\alpha}
 \def\b{\beta}
 \def\g{\gamma}
 \def\d{\delta}
 \def\e{\epsilon}

 \def\k{\kappa}

 \def\s{\sigma}
 \def\t{\tau}

 \def\G{\Gamma}
 \def\D{\Delta}

 \def\S{\Sigma}

 \def\o{\omega }
 
\def\equskip{\!\!\!\!\!\!\!\!} 
\def\la{\left\langle}
\def\ra{\right\rangle}
\def\raCL{\right\rangle_\text{CL}}
\def\raCLc{\right\rangle_\text{c,CL}}
\def\hf{\dfrac{1}{2}}
\def\action{\mathcal{S}}
\def\funcpart{\mathcal{Z}}
\def\Op{\mathcal{O}}

\def\bL{b}

\def\implies{\quad\Rightarrow\quad}

\def\tB{t^{(l)}}
\def\ttB{\t^{(l)}}

\begin{document}

\begin{titlepage}
\renewcommand{\thefootnote}{\fnsymbol{footnote}}
\flushright{\large KUNS-2352}
\vspace*{1cm}
    \begin{Large}
       \begin{center}
         {Bulk-boundary correlators in the hermitian matrix model and minimal Liouville gravity}
       \end{center}
    \end{Large}
\vspace{0.7cm}

\begin{center}
Jean-Emile B{\sc ourgine$^{1}$}\footnote
            {
e-mail address : 
jebourgine@sogang.ac.kr},
Goro I{\sc shiki$^{2,3}$}\footnote
            {
e-mail address : 
ishiki@post.kek.jp}
    {\sc and}
Chaiho R{\sc im$^{2}$}\footnote
           {
e-mail address : 
rimpine@sogang.ac.kr}\\
      
\vspace{0.7cm}                    
{\it Department of Physics$^{2}$ and 
  Center for Quantum Spacetime (CQUeST)$^{1,2}$
}\\
{\it Sogang University, Seoul 121-742, Korea}\\

\vspace{0.7cm}
{\it Department of Physics, Kyoto University,$^{3}$\\
Kyoto 606-8502, Japan}
\end{center}

\vspace{0.7cm}

\begin{abstract}
\noindent

We construct the one matrix model (MM) correlators corresponding to the general bulk-boundary correlation numbers of the minimal Liouville gravity (LG) on the disc. To find agreement between both discrete and continuous approach, we investigate the resonance transformation mixing boundary and bulk couplings. 
It leads to consider two sectors, depending on whether the matter part of the LG correlator is vanishing due to the fusion rules. In the vanishing case, we determine the explicit transformation of the boundary couplings at the first order in bulk couplings. In the non-vanishing case, no bulk-boundary resonance is involved and only the first order of pure boundary resonances have to be considered. Those are encoded in the matrix polynomials determined in our previous paper. We checked the agreement for the bulk-boundary correlators of MM and LG in several non-trivial cases. In this process, we developed an alternative method to derive the boundary resonance encoding polynomials.

\end{abstract}
\vfill
\end{titlepage}
\vfil\eject

\setcounter{footnote}{0}

\section{Introduction}
The correspondence between the two descriptions of 2D quantum gravity given by the matrix model (MM) and the Liouville gravity (LG) has been widely investigated for more than thirty years \cite{Ginsparg1993,DiFrancesco1995}. One of the key steps was the matching of the MM critical exponents with the gravitational dimension of the LG correlators \cite{Knizhnik1988,David1988,Distler1988}. Technically more elaborate, the matching between correlators was first performed in \cite{Moore1991} for the sphere one and two points functions. This correspondence involves the so-called resonance transformation, a finite renormalization of the couplings arising due to the ambiguity lying in the presence of contact terms. Since these early steps, various matrix models have been introduced (one and two hermitian matrix models, ADE, matrix chains, $O(n)$ loop gas model, q-Potts,...) to describe a discrete quantum gravity with different kind of matter fields. We concentrate here on the simplest one, the one hermitian matrix model which continuum description is provided by a minimal Liouville gravity, i.e. a quantum gravity which matter sector is given by a minimal model. In this context, the full resonance transformation on the sphere has been conjectured in \cite{Belavin2009}, and checked up to the fifth order \cite{Belavin2009,Tarnopolsky2010} (see also \cite{Zamolodchikov2005a} for the case of the gravitational scaling Lee-Yang model). Agreement between the two descriptions has also been verified on the disc with the trivial boundary conditions (BC) in \cite{Belavin2010}.

Recently, the correspondence between the one matrix model and minimal LG has been extended to worldsheets with arbitrary boundaries. More precisely, matrix correlators describing a disc with any boundary conditions have been constructed by two of the authors in \cite{Ishiki2010}. It was later shown \cite{Bourgine2010a} that this construction relies on a linear relation of decomposition satisfied by the FZZT branes \cite{Fateev2000,Teschner2000} of LG found in \cite{Seiberg2004}. In \cite{Bourgine2010b}, we constructed the matrix boundary operators introduced between two arbitrary boundary conditions. This construction involves the first order of a pure boundary resonance transformation that takes the form of a polynomial in the matrix.\footnote{Strictly speaking, this resonance transformation also includes the bulk LG cosmological constant linked to the dressed bulk identity operator. In ``pure boundary'', we understand here that no other bulk operators are involved except for this cosmological term of the critical action. All these statements will be clarified in the section three below.}

The purpose of this paper is to extend the previous constructions by considering matrix correlators describing a disc with an insertion of one bulk and one boundary operator. In the search for agreement between LG and MM correlators, we are led to distinguish two different cases. When the boundary operator Kac index does not belongs to the fusion rules of two copies of the bulk operator index, the matter part of the LG correlator has to vanish. This case involves a mixed bulk-boundary resonance transformation that will be determined at the first order by imposing the cancellation of MM correlators combinations. On the other hand, if the LG correlator is non-vanishing, no mixed bulk-boundary resonance arises for dimensional reasons. Thus, we only have to consider the pure boundary resonance already determined in \cite{Bourgine2010b}. We explicitly checked the agreement between MM and LG in a few non-trivial cases, confirming our general approach to the resonance transformation. In this process, we derived an alternative expression for the polynomials encoding the resonance transformation. This expression proved to be convenient when the Kac index of the boundary operators are small.

In the first section, we concentrate on the LG side and specialize the expression of the Liouville bulk-boundary correlator \cite{Hosomichi2001} to the case of a coupling with a degenerate matter operator. In section two, we explain how to construct the bulk-boundary matrix correlators and derive their expression in the continuum limit. The section three is the core of the paper, starting with general considerations on the resonance transformation and deriving the relation between MM and LG correlators. Then we show the agreement between the two approaches for both vanishing and non-vanishing LG correlators. Most of the technical details are assigned to the appendix.

\section{On the Liouville gravity side}
\subsection{Bulk-boundary correlator in Liouville gravity}
The Liouville gravity is the description of 2D quantum gravity with critical matter in the conformal gauge \cite{Polyakov1981}. The total action obeys the conformal symmetry and consists of three components, the Liouville, matter and ghost parts
The bulk Liouville action  $\action_L$ is given as 
\begin{equation}
\action_L=\dfrac{1}{4\pi}\int_\S{\sqrt{g}d^2z\left[g^{ab}\p_a\phi\p_b\phi+QR\phi+4\pi\mu e^{2\bL\phi}\right]},
\end{equation}
where $\mu$ is the bulk cosmological constant 
and $R$ the Ricci scalar associated to the fixed worldsheet metric $g^{ab}$. 
The background charge $Q$ is related to the Liouville parameter $\bL$ by $Q=\bL+1/\bL$,
which provides the central charge $c_L = 1+ 6Q^2$. 
The matter part is described by a conformal action with a central charge $c_M = 1 - 6q^2$
where $q=\bL-1/\bL$. 
The ghost part is the usual $(b,c)$ system of the bosonic string theory, with central charge $c_\text{gh}=-26$. 
The vanishing of the total conformal anomaly leads to a constraint on the central charges that fixes the Liouville parameter $\bL$.
In the following, we will concentrate on the $p$-critical models, realized as a LG with a matter sector given by a minimal model of the Lee-Yang series $(2,2p+1)$ for which $\bL^{-2}=p+1/2$.

For the critical gravity, the fields describing the conformal matter, metric and ghosts are formally decoupled. But in the correlation functions, any matter field needs to be dressed by the appropriate Liouville vertex operator to 
form a composite field of dimension $(1,1)$, $\Phi_{1,j}  e^{2\b_{1,j} \phi(z)} $. The dressing charge $\b_{1,j}$ is related to the bare operator scaling dimension through
\begin{equation}\label{dress_charge}
\b_{1j}=\dfrac{Q}{2}-P_{1j},\quad P_{1j}=\dfrac1{2\bL}-j\dfrac{\bL}{2} >0,
\end{equation}
where the indices $(1,j)$ denotes the position of $ \Phi_{1,j}$ in the Kac table, and $j=1,\cdots, p$. In order to respect the invariance under diffeomorphisms, the LG composite fields should be integrated over the worldsheet. However, the presence of conformal Killing vectors allow to fix the position of some of these operators, provided we multiply them with a suitable ghost factor. To fix the notation, the integrated (or fixed) dressed operator with Kac label $(1,p-J)$ will be denoted  $V_J$ with  $J$ running from zero to $p-1$.

On the disc, the Liouville action acquires a boundary term \cite{Fateev2000},
\begin{equation}
\action_L^{(B)}=\dfrac{1}{2\pi}\int_{\p\S}{g^{1/4}dx\ \left[QK\phi+\dfrac{2\pi}{\sqrt{\sin(\pi\bL^2)}}\mu_Be^{\bL\varphi}\right]},\quad \mu_B=\sqrt{\mu}\cosh\pi\bL s
\end{equation}
where $K$ is the intrinsic curvature of the boundary.\footnote{In order to simplify the relation between the LG and MM boundary cosmological constants, the factor $1/\sqrt{\sin(\pi\bL^2)}$ has been put outside the definition of $\mu_B$, contrary to the notation used in \cite{Fateev2000}.} At the quantum level, the boundary states are labeled by the parameter $s$, which is related to the boundary cosmological constant $\mu_B$ in the semi-classical limit by an hyperbolic cosine parameterization \cite{Teschner2000}. The Liouville gravity boundary states are tensor products of these states $|s>$ with a Cardy state $|(1,l)>$ describing the matter boundary conditions. Such states obey a linear decomposition property \cite{Seiberg2004},
\begin{equation}
|s,(1,l)>=\sum_{\a=-(l-1):2}^{l-1}{|s+i\a\bL,(1,1)>}
\end{equation}
where the sum runs from $-(l-1)$ to $l-1$ with steps of two. This decomposition has been shown to be consistent with the matrix model boundary construction on the disc \cite{Bourgine2010a}. However, its validity for topologies of higher genus may still need to be checked more carefully \cite{Kutasov2004,Atkin2011}.

Likewise in the bulk case, one may introduce dressed conformal matter boundary operators $\Phi^{B}_{1,j}  e^{\b_{1,j} \varphi(z)} $ with a dressing charge given by \ref{dress_charge}. The integrated dressed operators with Kac label $(1,2(l-k)-1)$ where $k$ is running from zero to $l-1$ will be denoted as $B_k$. For convenience, one may consider the dressed operators as a perturbation of the LG action by introducing external sources,
\begin{equation}
\Delta \action = \sum_{J=0}^{p-2} \t_J  V_J   +  \sum_{k=0}^{ \ell-2} \t^{(l)}_k  B_k  \,.
\end{equation}
The symmetry under translation of the Liouville field leads to associate a
gravitational dimension to the coupling constants. Assuming a dimension two for the bulk cosmological constant $\mu$, the other coupling dimensions are found to be $[\t_J]=p+1-J$ and $[\t^{(l)}_k]=l-k$. It is noted that by definition, the bulk and boundary cosmological constants couple to the dressed identity operators, $\t_{p-1}=\mu$ and $\t^{(l)}_{l-1}=\mu_B$, $[\mu_B]=1$. The sphere and disc partition functions respectively scale as $2Q/\bL$ and $Q/\bL$.

The bulk-boundary correlation number we are considering can be written as 
\begin{equation}\label{corr_LG}
\la V_J B_k\ra_\text{LG}^{(l)}=\mu^{(J+k-l+1/2)/2}\ \mathcal{N}_{Jk}^{(l)}~~R(P_J,\b_k,s)
\end{equation}
with $P_J \equiv P_{1,p-J}=\frac12(J+1/2)\bL$ and $\b_k \equiv \b_{1,2(l-k)-1}=(l-k)\bL$. Because of the presence of conformal Killing vectors,
one may simply fix the position of both boundary and bulk operators, avoiding this integration over coordinates. The coefficient $\mathcal{N}_{Jk}^{(l)}$ is independent of the Liouville parameter $s$ and $\mu$, it takes into account the matter and ghost factors, as well as some additional multiplicative constant in the expression of the Liouville correlator.\footnote{The exact expression of the factor $\mathcal{N}_{Jk}^{(l)}$ depends on the proper normalization of the operators. This difference of normalization between MM and LG correlators is sometimes referred as ``leg factors'', they corresponds to a multiplicative degree of freedom for the coupling constants. Here, it will be fixed by requiring monic polynomials in the resonance transformations of the section three.} This coefficient vanishes when the boundary operator Kac index does not  belong to the fusion algebra of the modules associated to two copies of the bulk operator, i.e. 
\begin{equation}\label{prop_N}
\mathcal{N}_{Jk}^{(l)}=0\quad\text{when}\quad l-k>p-J\,.
\end{equation}
In the section \refOld{sec_nvan_corr} below, we will compare only the $\mu$- and $s$- dependent part of MM and LG correlators. In this manner, we do not need the explicit expression for the coefficients $\mathcal{N}_{Jk}^{(l)}$ and the property \ref{prop_N} will be sufficient for our purpose.

The $s$-dependent factor in \ref{corr_LG} has been obtained in Liouville field theory 
and is given in \cite{Hosomichi2001} as a Fourier transform
\begin{equation}\label{four_trans}
R(P, \beta, s) =-i\int_\mathcal{C} d\s e^{-2\pi s\s}
\tilde{R}(P, \beta, \s) ,\quad \tilde{R}(P, \beta, \s)= \frac{S(\delta_++\beta)S(\delta_-+\beta)}{S(\delta_++Q)S(\delta_-+Q)}
\end{equation}
where $\delta_\pm=\s-\frac12\b\mp P$ and $S(x)$ denotes the double sine function \cite{Fateev2000} which satisfies the shift properties
\begin{equation}\label{property of S}
S(x+\bL)=2\sin(\pi \bL x)S(x),\quad S(x+1/\bL)=2\sin(\pi x/\bL)S(x),\quad S(x)S(Q-x)=1.
\end{equation}
In the context of Liouville theory, the momenta $P$ related to the charge of vertex operator by $\b=Q/2-P$ are purely imaginary. In this case, the Fourier transform in \cite{Hosomichi2001} is well defined and the contour $\mathcal{C}$ in the integral \ref{four_trans} is along the imaginary axis. However, in the context of Liouville gravity, these momenta are real and the convergence of the Fourier transform is no longer guaranteed. As a consequence, the contour $\mathcal{C}$ should be appropriately deformed (see details in the appendix \refOld{appA}). 

One may notice that $R(P, \b, s) $  is an even function of $P$ and $s$;
\begin{equation}
R(P, \b, s)= R(- P, \b, s)\,, \quad
R(P, \b, s)= R( P, \b, -s) \,.
\end{equation} 
For the degenerate values $\b=k\bL$, $k\in\mathbb{Z}^+$, 
$\tilde R $ is reduced to a ratio of sines due to the shift properties satisfied by $S$,
\begin{align}
\tilde{R}(P_J, k\bL, \s)
= 4^{k-2}
\frac{\prod_{\a=1}^{k-1}
\sin(\pi b \delta_+ +\pi \a \bL^2) \sin(\pi b \delta_-+ \pi \a \bL^2)}
{\sin(\pi \delta_+ /\bL) \sin(\pi \delta_- /\bL)}.
\label{sin_left}
\end{align}
The contour integration can be performed, leading to an infinite sum over residues,
\begin{align}
\begin{split}\label{expr_R}
R(P_J, k\bL, s)=\dfrac{4^{k-1}\bL}{\sin(2\pi P_J/\bL)}\Bigg(
&\sum_{n=0}^{\infty} \prod_{\a=1}^{k-1}
\sin\{\pi \bL^2(\a+n)\} \sin \{\pi \bL^2(\a+n)+2\pi P_J\bL\}
e^{-\pi\bL s (2n +k +2P_J/\bL)}\\
-&\sum_{n=0}^{\infty} \prod_{\a=1}^{k-1}
\sin\{\pi \bL^2(\a+n)\} \sin \{\pi \bL^2(\a+n)-2\pi P_J \bL\}
e^{-\pi\bL s (2n +k-2P_J/\bL)}\Bigg).
\end{split}
\end{align}

In the simple case $k=1$, the summation can be easily done and we get the simple expression
\begin{equation}\label{start_pt}
R(P_J,\bL,s)=-\dfrac{\bL}{\sinh(2\pi P_J/\bL)}\dfrac{\sinh{2\pi P_J s}}{\sinh\pi\bL s}.
\end{equation}
Unfortunately, the formula \ref{expr_R} is not convenient for comparison with the matrix model results. This is why in the next section we shall exploit the recursion relations over $s$ satisfied by $R$ in order to propose an alternative expression.

\subsection{Shift identities}
The bulk-boundary correlation $R(P_J, k\bL, s)$ obeys several shift relations that are supposed to be related to the bulk \cite{Witten1992,Kutasov1992} and boundary \cite{Kostov2004,Kostov2004a} ground ring structures. They provide a nice recursion relation which bypass  the complicated summed expression and will turned to be useful below to relate MM and LG correlators. The readers who are not interested in the technical details might  skip this subsection.

The simplest relation is obtained from the expression \ref{expr_R}, when $s$ is shifted by  $\pm i/\bL$; 
\begin{equation}\label{LG_shift_1bL}
R(P_J,k\bL,s+i/\bL)+R(P_J,k\bL,s-i/\bL)=(-1)^{k} 2\cos(2\pi P_J/\bL)\ R(P_J,k\bL,s).
\end{equation}
Another relation can be derived using the properties of the double sine function to show that
\begin{equation}
\tilde{R}(P_J+\bL/2,(k+1)\bL,\s)=2(\cos(2\pi\bL \s)-\cos2\pi\bL^2(k+P_J/\bL))\ \tilde{R}(P_J,k\bL,\s).
\end{equation}
Plugging this into the Fourier transform, the first cosine just gives a sum over shifts of $s$, and we end up with
\begin{equation}\label{bulk_rec}
R(P_J\pm\bL/2,(k+1)\bL,s)=R(P_J,k\bL,s+i\bL)+R(P_J,k\bL,s-i\bL)-2\cos2\pi\bL^2(k\pm P_J/\bL)\ R(P_J,k\bL,s)
\end{equation}
where a similar relation was also derived for negative shifts.

The most interesting relation is obtained if by shifting $s$ and $k$ but not $P_J$. 
The shift relation is properly derived in the appendix \ref{appB} and  it reads
\begin{align} 
\cosh\pi\bL s\ R(P_J,(k+1)\bL,s) &=2\cos (2\pi\bL P_J) ~ R(P_J,k\bL,s)
-  c_k \left[R(P_J,k\bL,s+i\bL)+R(P_J,k\bL,s-i\bL)\right]
\nonumber\\
\sinh\pi\bL s\ R(P_J,(k+1)\bL,s)&=i s_k ~ \left[R(P_J,k\bL,s+i\bL)-R(P_J,k\bL,s-i\bL)\right],
\label{rec_LG}
\end{align}
or equivalently
\begin{equation}
R(P_J,k\bL,s\pm i\bL)=\dfrac{\cos 2P_J\pi\bL}{c_k}R(P_J,k\bL,s)-\hf\left(\dfrac{\cosh\pi\bL s}{c_k}\pm i\dfrac{\sinh\pi\bL s}{s_k}\right)R(P_J,(k+1)\bL,s),
\end{equation}
where we used the shortcut notation
\begin{equation}\label{shortcut}
c_k=\cos\pi\bL^2 k,\quad s_k=\sin\pi\bL^2 k\,.
\end{equation}
Together with the knowledge of the expression \ref{start_pt} for the operator $B_1$, this last relation fully determines $R$ at any $\b_k$. Indeed, it is easy to show that the following expression satisfies recursively the second shift relation in \ref{rec_LG} with the initial condition given by \ref{start_pt} (see appendix \refOld{appC}),
\begin{equation}\label{sol_bb_LG}
R(P_J,k\bL,s)=-u^{k-1}\dfrac{\bL2^{k-1}}{\sinh(2\pi P_J/\bL)}\prod_{\g=1}^{k-1}{\sinh^2 i\pi\bL^2\g}\ \sum_{\a=-(k-1):2}^{k-1}{\prod_{\superp{\b=-(k-1):2}{\b\neq\a}}^{k-1}{(x_\a-x_\b)^{-1}}\ \dfrac{\sinh2\pi P_Js_\a}{\sinh\pi\bL s_\a}}
\end{equation}
with $s_\a=s+i\a\bL$ and $x_\a=x(s_\a)=u\cosh\pi\bL s_\a$. The expression for $k=2$ can be simplified into
\begin{equation}\label{expr_P2}
R(P_J,2\bL,s)=-\dfrac{\bL s_1}{\sinh(2\pi P_J/\bL)}\dfrac{c_r^2-c_1^2}{\cosh^2\pi\bL s-c_1^2}\sum_\pm{\pm\dfrac{\sinh(2P_J\pm\bL)\pi s}{s_{r\pm1}\ \sinh\pi\bL s}}.
\end{equation}
Finally, a recursion relation on $k$ can be derived by considering $R(P_J,k\bL,s+i\bL-i\bL)$,
\begin{equation}\label{rec_LG_b}
2\dfrac{c_{k+1}s_k}{s_{2k+1}}\left(c_k^2-c_r^2\right)R(P_J,k\bL,s)=-c_r\cosh\pi\bL s\ R(P_J,(k+1)\bL,s)+\dfrac{c_k(\cosh^2\pi\bL s-c_{k+1}^2)}{2s_{2k+1}s_{k+1}}R(P_J,(k+2)\bL,s),
\end{equation}
with $ r=2P_J/\bL$ in $c_r$.

\section{On the matrix side}\label{sec_MM}
The one matrix model partition function is given by an integral\footnote{This integral should be understood here as a formal series obtained by expanding the exponential, keeping only the quadratic term, the coefficient of which must be negative.} over an $N\times N$ hermitian matrix $M$,
\begin{equation}
\funcpart_\text{1MM}=\int{dM e^{-\frac{N}{g}\tr V(M)}}
\end{equation}
where $V(M)$ is a polynomial potential chosen to achieve the $(p+1)$-th Kazakov multi-critical point \cite{Kazakov1989} (for a review of this model, see \cite{Ginsparg1993,DiFrancesco1995}). The partition function and the correlators can be expanded in powers of $N^{-2}$, each term being associated to a different topology. We focus here on the first term of the series which describe the planar topologies (sphere or disc). These first order terms depend on the t'Hooft parameter $\k^2=1/g$ which weight the number of vertices of the planar Feynman diagrams. Since the parameter $\k$ controls the area of the discrete surfaces dual to the Feynman diagrams, it is sometimes referred as the bare cosmological constant. In the continuum limit, $\k$ is sent to a critical value $\k^*$ where the mean area diverge, and we define the renormalized coupling  $\e^{1/\bL^2}t_0=\k-\k^*$ where $\e\sim N^{-2/2p+3}$ is the lattice size cut-off. To this coupling is assigned the dimension $[t_0]=p+1$.

In order to compare the $(p+1)$-th multi-critical MM with the $(2,2p+1)$ minimal Liouville gravity, we first need to introduce the KdV deformations \cite{Gross1990,Banks1989,Douglas1989}. These linear deformations of the potential $V(M)$ by the other multi-critical potentials $V(M)\to V(M)+\sum_J{\bar{t}_J^{(\text{KdV})}V_J^{(\text{KdV})}}$ lead to the string equation that determines the string susceptibility - or two-punctured sphere partition function - $u$ as a function of the deformations. Under a suitable normalization, the string equation reads
\begin{equation}
u^{p+1}-\sum_{J=0}^{p-1}{t_J^{(\text{KdV})}u^J}=0.
\end{equation}
where $t_J^{(\text{KdV})}$ is the renormalized $\bar{t}_J^{(\text{KdV})}$. The $(p+1)$-th multi-critical point, denoted $\ast$, is defined as the point where all the perturbations, including the MM cosmological constant
$t_0=t_0^{(\text{KdV})}$, are turned off, except for the coupling $t_{p-1}^{(\text{KdV})}$ of dimension two. At this point, the MM can be compared to LG and $t_{p-1}^{(\text{KdV})}$ corresponds to the LG cosmological constant $\mu$.\footnote{Contrary to the unitary models (e.g. the $O(n)$ matrix model \cite{Kostov2008}), here the area is no longer measured by the dressed identity operators, but by the operator of highest dimension $V_0$. In this sense, the MM renormalized cosmological constant $t_0^{(\text{KdV})}$ does not corresponds to the LG cosmological constant $t_{p-1}^{(\text{KdV})}=\mu$. To avoid confusion, in the following we should only refer to $\mu$ under the denomination ``bulk cosmological constant''.} In particular, $u|_\ast=\sqrt{\mu}$ and the resolvent defined below is identified with the LG identity boundary 1pt function on the disc.

As mentioned in the introduction, the bulk resonance transformation has already been treated in \cite{Belavin2009}. At the level of insertion of a single bulk operator, this unnecessary complication can be avoided by a linear redefinition of the polynomials that perturb the matrix potential. These redefined potentials, denoted here $V_J$, absorb the first order (but all orders in $\mu$) of the resonance transformation. After this redefinition, the new potentials couple to bare parameters $\bar{t}_J$ that can be directly identified with the parameters $\tau_J$ perturbing the minimal LG in the continuum limit. In this setting, the bulk 1-pt functions are obtained from the matrix model correlators as,
\begin{equation}\label{CFT_frame}
\la\tr V_J(M)\raCL=\left.\dfrac{\p \mathcal{Z}_\text{1MM}}{\p t_J}\right\vert_\ast=\la V_J\ra_\text{LG},
\end{equation}
where the bulk couplings are turned off at $\ast$, except for $t_{p-1}=\mu$. With this choice, all the MM bulk 1-pt functions vanishes apart from the one with $J=p-1$.

Let us turn to the boundary effect. The disc with one marked point and trivial boundary conditions (i.e. leading to $(1,1)$ matter BC in the continuum limit) is given by the resolvent, defined as
\begin{equation}\label{expr_res}
 W(\bar{x})=\lim_{N\to\infty}\dfrac{1}{N}\la\tr\dfrac1{\bar{x}-M}\ra=\hf V'(\bar{x})+\o(\bar{x}).
\end{equation}
This quantity is the Stielges transform of the eigenvalue density for the matrix $M$.
 The bare boundary cosmological constant $\bar{x}$ controls the length of the boundary of the discretized surfaces. In the continuum limit, it is also sent to a critical value $\bar{x}^*$ where the mean boundary length diverges. This critical value can be taken to be zero by a shift of the matrix, and we define the renormalized boundary cosmological constant as $\e x=\bar{x}$, $[x]=1$. In the process, we have to throw away the non-universal contributions which are polynomials in $\bar{x}$ such as the term $V'(\bar{x})$ appearing in the expression \ref{expr_res} of the resolvent. Since in the following we focus on the expression of the correlators in this continuum limit, we automatically throw away all the non-critical terms and write directly \cite{Daul1993}:
\begin{equation}
\o(x)=\la\tr\dfrac1{x-M}\ra_\text{CL}=u^{1/\bL^2}\cosh(\pi s/\bL),\quad x=u\cosh\pi\bL s.
\end{equation}
The uniformizing parameterization of $x$ has been introduced in order to resolve the branch cut over $]-\infty,-u]$ of the resolvent. The right bound of the branch cut is identified with the ``string susceptibility'' $u$.

On the disc topology, the boundary describing a matter with $(1,l)$ BC have been constructed in \cite{Ishiki2010}
by allowing the matrix to additionally interact with Gaussian vectors with flavors.
For instance, the one and two-point functions of minimal LG with matter BC $(1,l)$ 
and cosmological constants $x(s)$  (two identical boundaries for the two-point function) have been
identified in \cite{Bourgine2010b} as
\begin{equation}\label{mat_12pt}
\la B_k\ra_\text{LG}^{(l)}=\la\tr\dfrac{M^{\ell-k}}{F_l(x,M)}\ra_\text{CL},\quad \la B_kB_{m}\ra_\text{LG}^{(l)}=\la\tr\dfrac{P^{(l),k}(x,M)P^{(l),m}(x,M)}{F_l(x,M)^2}\raCL
\end{equation}
where the leg factors have been included in the definition of the LG correlators
and 
\begin{equation}
F_l(x,M)=\prod_{\a=-(l-1):2}^{l-1}{(M-\bar{x}_\a)},\quad x_\a=u\cosh\pi\bL s_\a,\quad s_\a=s+i\a\bL
\end{equation}
creates the corresponding MM boundary.
The polynomials $P^{(l),k}(x,M)$, determined in \cite{Bourgine2010b}, 
encode the pure boundary resonance transformation 
and will be defined more precisely in the section \refOld{sec_res_trans} below.

The MM correlators with identical boundaries $F_l(x,M)$ can be derived from the following disc partition function 
in the presence of boundary sources $t_a^{(l)} M^a$,
\begin{equation}
\label{bdry_parttition} 
 \funcpart^{(l)}(t_a^{(l)})=\la\tr\log\left(F_l(x,M)+\sum_{a=0}^{l-1}{t_a^{(l)}M^a}\right)\ra_\text{CL}\,.
\end{equation}
To investigate the bulk-boundary matrix correlator, we need to consider the disc partition function of Eq.~\ref{bdry_parttition} in the presence of the bulk sources in the CFT frame, as in \ref{CFT_frame}. The matrix correlator we are interested in is defined as
\begin{equation}\label{def_ojkl}
\Op_{Ja}^{(l)}(x)=\left.\dfrac{\p^2\funcpart^{(l)}}{\p t_J\p\tB_a}\right\vert_\ast=\la\tr\dfrac{M^a}{F_l(x,M)}\tr V_J(M)\ra_\text{c,CL}
\end{equation}
where 'c' denotes the connected part  and $\ast$ all the non-trivial couplings turned off. 
This quantity is easily evaluated from the expansion of the product $F_l$ (see identity (A.1) of \cite{Bourgine2010a}),
\begin{equation}
\Op_{Ja}^{(l)}(x)=-\sum_{\a=-(l-1):2}^{l-1}{\prod_{\superp{\b=-(l-1):2}{\b\neq\a}}^{l-1}(x_\a-x_\b)^{-1}\la\tr\dfrac{M^a}{x_\a-M}\tr V_J(M)\ra_\text{c,CL}}.
\end{equation}
We recursively make use of the property 
\begin{equation}\label{pow_van}
\sum_{\a=-(l-1):2}^{l-1}{\prod_{\superp{\b=-(l-1):2}{\b\neq\a}}^{l-1}(x_\a-x_\b)^{-1}x_\a^a}=0,\quad \text{for }a<l-1,
\end{equation}
in order to rewrite the matrix correlators as
\begin{equation}
\Op_{Ja}^{(l)}(x)=-\sum_{\a=-(l-1):2}^{l-1}{\prod_{\superp{\b=-(l-1):2}{\b\neq\a}}^{l-1}(x_\a-x_\b)^{-1}x_\a^a\la\tr\dfrac{1}{x_\a-M}\tr V_J(M)\ra_\text{c,CL}}
\end{equation}
This decomposition involves the derivative of the known bulk 1-pt functions with respect to the boundary cosmological constant \cite{Belavin2010},
\begin{equation}
\Op_{J0}^{(1)}(x)=-\la\tr\dfrac{1}{x-M}\tr V_J(M)\ra_\text{c,CL}=\hf\bL^{-2}u^{2P_J/\bL-1}\dfrac{\sinh2\pi P_J s}{\sinh\pi\bL s}
\end{equation}
where the normalization has been fixed such that
\begin{equation}
\Op_{p-1\ 0}^{(1)}(x)=-\p_\mu\vert_\text{$x$ fixed}\o=\frac12\bL^{-2}u^{p-3/2}\dfrac{\sinh\pi s(1/\bL-\bL)}{\sinh\pi\bL s}.
\end{equation}
We end up with the following expression for the bulk-boundary matrix correlators,
\begin{equation}\label{expr_ojkl}
\Op_{Ja}^{(l)}(x)=\hf\bL^{-2}u^{2P_J/\bL-1}\sum_{\a=-(l-1)}^{l-1}{\prod_{\superp{\b=-(l-1)}{\b\neq\a}}^{l-1}(x_\a-x_\b)^{-1}x_\a^a\dfrac{\sinh(2\pi P_J s_\a)}{\sinh(\pi\bL s_\a)}}.
\end{equation}
It should be noted that the $s$-dependent part of 
$\Op_{J 0}^{(l)}(x)$ exactly reproduces
the correlator of LG, 
$R(P_J,k\bL,s)$ in Eq.~\ref{sol_bb_LG}. 

Finally, let us stress that the expression \ref{expr_ojkl} satisfies shift relations similar to those obtained for the LG correlator in section 1.2.
\begin{align}
\begin{split}\label{equ_shift_MM}
&\Op_{Ja}^{(l)}(s+i/\bL)+\Op_{Ja}^{(l)}(s-i/\bL)=2(-1)^{l-a}\cosh(2\pi P_J/\bL)\ \Op_{Ja}^{(l)}(s),\\
&2\Op_{J\ a+1}^{(l)}(s)=\Op_{J+1\ a}^{(l)}(s)+\Op_{J-1\ a}^{(l)}(s),\\
&\Op_{Ja}^{(l)}(s\pm i\bL)=\Op_{J\ a+1}^{(l+1)}(s)-x_{\mp l}\Op_{Ja}^{(l+1)}(s).
\end{split}
\end{align}
The third identity is derived from a relation that can be linked to the insertion of a boundary ground ring operator, as explained in the appendix of \cite{Bourgine2010b}.

\section{Comparison of matrix models with Liouville Gravity}
\subsection{The resonance transformation}\label{sec_res_trans}
In order to compare the matrix model with the LG results, we have to take into account a possible redefinition of the coupling constants known as the resonance transformation. This transformation is a consequence of the ambiguity in the definition of the so-called contact terms, i.e. the values of the correlators that contain operators taken at coinciding points. Such contact terms can be reabsorbed into a finite renormalization of the couplings. Coinciding bulk operators lead to pure bulk coupling resonance. This phenomenon was first introduced in the matrix model context in \cite{Moore1991} and then later investigated for the perturbed sphere partition function in \cite{Belavin2009}. When dealing with a worldsheet having a boundary, we have to take into account boundary operators at coinciding points which leads to pure boundary couplings resonance, i.e. redefinition of the boundary couplings involving only other boundary couplings. In addition, we also have to take into account a bulk operator coinciding with the boundary (or equivalently with its mirror image) and bulk operators coinciding with boundary ones. These two phenomena lead to a bulk-boundary resonance which translates into the presence of bulk couplings in the resonance transformation of the boundary ones. It is stressed that no boundary couplings can be involved in the resonance of the bulk ones. The resonance transformation can be seen as a finite renormalization and this should be related to the study made in \cite{Fredenhagen2006}, where it was noticed that no boundary couplings arise in the RG equations of bulk couplings. On the contrary bulk couplings in the RG equations of boundary ones generate induced boundary flows. 

In the problem we are considering, we have already taken into account the bulk resonance transformation in a suitable definition of the matrix potential deformations $V_J$, and the general boundary resonance transformation writes
\begin{equation}\label{res}
\tB_a=\t^{(l)}_a+\sum_{\rho,\nu}c_{\rho,\nu}\prod_{J,k}{\t_J^{\rho_J}\t^{(l)\ {\nu_k}}_k},\quad [\tB_a]=\sum_{J}\rho_J[\t_J]+\sum_k\nu_k[\t^{(l)}_k],
\end{equation}
where the $c_{\rho,\nu}$ are just numerical constants. The resonance condition on the coupling constants dimension that restricts the form of the RHS gave its name to the transformation. We recall the dimensions of the couplings: $[\t_J]=p+1-J$ and $[\t_k^{(l)}]=l-k$.

The bulk and boundary cosmological constants $\mu=\t_{p-1}$ and $\mu_B=\t^{(l)}_{l-1}$ are not turned off at the point $\ast$ where the other perturbations vanishes. These two couplings are of minimal dimension among respectively the bulk and boundary ones, so that no other couplings can appear in their resonance transformation: $t_{l-1}=\mu$ and $\tB_{l-1}=-\frac{s_l}{s_1}\mu_B$ with $s_k=\sin\pi\bL^2 k$ (or $x=\mu_B$, the coefficient $-s_l/s_1$ appearing in the normalization of $\tB_{l-1}$ is due to the fact that the monomial of degree $l-1$ in $F_l(x,M)$ is proportional to $-\frac{s_l}{s_1}x$, see \cite{Ishiki2010,Bourgine2010a}). The boundary resonance transformation relevant for the study of boundary one and two points functions, as well as bulk-boundary correlators, writes
\begin{equation}\label{res_trans}
\tB_a=P_a^{(l)}(\mu,\mu_B)+\sum_{k=a}^{l-2}P_a^{(l),k}(\mu,\mu_B)\t_k^{(l)}+\sum_{k,m=a}^{l-2}P_a^{(l),km}(\mu,\mu_B)\t_k^{(l)}\t_m^{(l)}+\sum_{J=p+1-l+a}^{p-2}Q_a^{(l),J}(\mu,\mu_B)\t_J,
\end{equation}
The function $P_a^{\cdots}$ and $Q_a^{\cdots}$ are polynomials in $\mu$ and $\mu_B$ and the summations are restricted because of the resonance condition on the gravitational dimensions. Multiplication of the boundary couplings by a constant will only change the normalization of the boundary operator, leading to a different leg factor. In order to fix this degree of freedom, we assume here that at the first order $t_a^{(l)}=\t_a^{(l)}$, or $P_a^{(l),a}=1$ .

To understand how the transformation \ref{res_trans} relates the correlators of the two theories, we first consider the LG boundary 1-pt function with a non-trivial operator inserted ($B_k\neq B_{l-1}$) and apply the chain rule,
\begin{equation}
 \la B_k\ra_\text{LG}^{(l)}=\dfrac{\p\funcpart^{(l)}}{\p \t^{(l)}_k}=\sum_{a=0}^{k}\left.\dfrac{\p\tB_a}{\p \t^{(l)}_k}\dfrac{\p\funcpart^{(l)}}{\p \tB_a}\right\vert_\ast=\sum_{a=0}^{k}P_a^{(l),k}\la\tr\dfrac{M^a}{F_l(M)}\raCL=0
\end{equation}
since the matrix correlators with powers of $M^a$, $a<l-1$ are vanishing due to \ref{pow_van}. In a similar way for the identity operator we get
\begin{equation}
\la B_{l-1}\ra_\text{LG}^{(l)}=\dfrac{\p\funcpart^{(l)}}{\p \t^{(l)}_{l-1}}=\sum_{a=0}^{l-1}\left.\dfrac{\p\tB_a}{\p \t^{(l)}_{l-1}}\dfrac{\p\funcpart^{(l)}}{\p \tB_a}\right\vert_\ast=\sum_{a=0}^{l-1}\p_{\mu_B}P_a^{(l)}\la\tr\dfrac{M^a}{F_l(M)}\raCL=(-1)^{l}\o(x)\p_{\mu_B}P_{l-1}^{(l)}
\end{equation}
This is in indeed what we observe \cite{Bourgine2010a} since $P_{l-1}^{(l)}(\mu,\mu_B)=-(s_l/s_1)\mu_B$.

We now turn to the boundary 2pt functions and start again with the non-trivial operators $k\neq l-1\neq m$,
\begin{align}
 \begin{split}\label{45}
 \la B_kB_m\ra_\text{LG}^{(l)}&=\left.\dfrac{\p}{\p\t_k^{(l)}}\left(\sum_{a=0}^m\dfrac{\p\tB_a}{\p\t_m^{(l)}}\dfrac{\p\funcpart^{(l)}}{\tB_a}\right)\right\vert_\ast\\
&=\sum_{a=0}^{\inf(k,m)}\left.\dfrac{\p^2\tB_a}{\p\ttB_k\p\ttB_m}\dfrac{\p\funcpart^{(l)}}{\p\tB_a}\right\vert_\ast+\sum_{a=0}^m\sum_{b=0}^k\left.\dfrac{\p\tB_a}{\p\ttB_m}\dfrac{\p\tB_b}{\p\ttB_k}\dfrac{\p^2\funcpart^{(l)}}{\p\tB_a\p\tB_b}\right\vert_\ast\\
&=\sum_{a=0}^{\inf(k,m)}\dfrac{\p^2\tB_a}{\p\ttB_k\p\ttB_m}\la\tr\dfrac{M^a}{F_l(x,M)}\raCL-\sum_{a=0}^m\sum_{b=0}^k\dfrac{\p\tB_a}{\p\ttB_m}\dfrac{\p\tB_b}{\p\ttB_k}\la\tr\dfrac{M^{a+b}}{F_l(x,M)^2}\raCL.\\
 \end{split}
\end{align}
As in the case of the boundary 1pt functions, the first sum vanishes so that the quadratic term in the resonance transformation \ref{res} plays no role here. The second term defines the two polynomials that were determined in \cite{Bourgine2010b},
\begin{equation}
\la B_kB_m\ra_\text{LG}^{(l)}=-\la\tr\dfrac{P^{(l),k}(x,M)P^{(l),m}(x,M)}{F_l(x,M)^2}\raCL,\quad P^{(l),k}(x,M)=\sum_{a=0}^k{P_a^{(l),k}(\mu,x)M^a}.
\end{equation}
Since we imposed $P_k^{(l),k}=1$, $P^{(l),k}$ is a monic polynomial in $M$. When one of the operator is trivial ($m=l-1$), the computation is similar to \ref{45} but we have to define the matrix polynomial separatly as
\begin{equation}
P^{(l),l-1}(x,M)=-\dfrac{s_1}{s_l}\sum_{a=0}^{l-1}{\p_{x}P_a^{(l)}(\mu,x)M^a}.
\end{equation}
where the normalization has been fixed in order to have again a monic polynomial in $M$. In the following we will not specify anymore whether or not the operators are different from the identity, but just keep in mind this subtlety in the definition of $P^{(l),l-1}$.

Finally, let us investigate the case of interest for this paper, the bulk-boundary correlators
\begin{equation}
\label{interm}
\la V_JB_k\ra_\text{LG}^{(l)}=\left.\dfrac{\p}{\p\t_J}\left[\sum_{a=0}^k\dfrac{\p\tB_a}{\p\t^{(l)}_k}\dfrac{\p\funcpart^{(l)}}{\p\tB_a}\right]\right\vert_\ast=\sum_{a=0}^k\left.\dfrac{\p\tB_a}{\p\t^{(l)}_k}\dfrac{\p^2\funcpart^{(l)}}{\p \t_J\p\tB_a}\right\vert_\ast+\sum_{a=0}^k\left.\dfrac{\p^2\tB_a}{\p\t_J\p\t^{(l)}_k}\dfrac{\p\funcpart^{(l)}}{\p\tB_a}\right\vert_\ast.
\end{equation}
As in the case of the boundary 1pt and 2pt functions, the second term vanishes because either the correlator is zero ($a<l-1$) or $a=l-1$ and the coupling $t_{l-1}^{(l)}$ has no resonance involving $\t_J$. The first term in \ref{interm} is a sum of two contributions since the variables $t_a^{(l)}$ may depends on $\t_J$,
\begin{equation}
\la V_JB_k\ra_\text{LG}^{(l)}=\sum_{a=0}^k\left.\dfrac{\p\tB_a}{\p\t^{(l)}_k}\dfrac{\p^2\funcpart^{(l)}}{\p t_J\p\tB_a}\right\vert_\ast+\sum_{a=0}^k\sum_{b=0}^{l+J-p-1}\left.\dfrac{\p\tB_a}{\p\t^{(l)}_k}\dfrac{\p\tB_b}{\p\t_J}\dfrac{\p^2\funcpart^{(l)}}{\p\tB_a\p\tB_b}\right\vert_\ast
\end{equation}
Introducing the previous polynomials $P^{(l),k}$ and the bulk-boundary encoding polynomials\footnote{Note that since the relative normalization of the bulk and boundary couplings have already been fixed, there is no reason to require that $Q^{(l),J}$ is a monic polynomial in $M$.}
\begin{equation}
Q^{(l),J}(x,M)=\sum_{b=0}^{l+J-p-1}{Q^{(l),J}_b(x)M^b},\quad Q^{(l),p-1}(x,M)=\sum_{b=0}^{l-2}{\p_\mu P^{(l)}_b(x)M^b}
\end{equation}
we finally get
\begin{equation}\label{rel_res}
\la V_JB_k\ra_\text{LG}^{(l)}=\la\tr\dfrac{P^{(l),k}(x,M)}{F_l(x,M)}\tr V_J(M)\ra_\text{c,CL}-\la\tr\dfrac{P^{(l),k}(x,M)Q^{(l),J}(x,M)}{F_l(x,M)^2}\raCL
\end{equation}
The gravitational dimension of the polynomials can be deduced from this relation,
\begin{equation}
[\la V_JB_k\ra]=k+J-l+1/2,\quad [\Op_{Ja}^{(l)}]=a+J-l+1/2\quad \implies\quad [P^{(l),k}]=k,\quad [Q^{(l),J}]=J+l-p-1.
\end{equation}

The formula \ref{rel_res} implies that we have to consider two different cases. When the LG correlator vanishes, both terms in \ref{rel_res} contribute. It appears that imposing the vanishing of the combination of the two MM correlators provides a sufficient number of constraints to fully determine the bulk-boundary resonance term $Q^{(l),J}$. It will be done in the section \refOld{sec_van_corr}. 

On the other hand, when the LG correlator is non-zero, i.e. $l-k\leq p-J$, we observe that the term which involves $Q^{(l),J}$ is not present in \ref{rel_res}. Indeed, in this case the boundary operator $B_k$ does not belong to the fusion outcome of the two copies of the bulk operators involved in the resonance of $t_{b<l+J-p-1}^{(l)}$. Because of the orthogonality property of the boundary 2pt functions, there can be no contribution from a contact term arising when a bulk operator meets the boundary. We will explicitly check this result by relating MM and LG correlators for several non-trivial cases ($k=0,1,2,l-2,l-1$) in the section \refOld{sec_nvan_corr} below.

\subsection{Preliminary checks}
As the first verification of the formula derived for both MM and LG correlator, we check the agreement of both expressions when no resonance is involved. This is the case for the boundary operator $B_0$ since the boundary coupling $\t_0^{(l)}$ only enters the resonance transformation of $t_0^{(l)}$ in \ref{res_trans} and $P^{(l),0}$ simply equals to one. For the non-vanishing case $l\leq p-J$ where $Q^{(l),J}$ is not involved, the expression \ref{expr_ojkl} of $\Op_{J0}^{(l)}(s)$ reproduces \ref{corr_LG} for the LG correlator $\la V_JB_0\ra_\text{LG}^{(l)}$,
\begin{equation}\label{k0}
\Op_{J0}^{(l)}(s)=u^{J-l+1/2}\mathcal{N}_{J0}^{(l)}\ R(P_J,l\bL,s),\quad \mathcal{N}_{J0}^{(l)}=-\bL^{-3}2^{-l}\dfrac{\sinh(2\pi P_J/\bL)}{\prod_{\g=1}^{l-1}{\sinh^2 i\pi\bL^2\g}}.
\end{equation}

As the second verification of the previous formalism, we may investigate the shift relations involving $s\to s\pm i/\bL$. More precisely, we should check that the sign flips arising due to such shifts (see \ref{LG_shift_1bL} and \ref{equ_shift_MM}) are compatible with the formula \ref{rel_res} and the definition of the polynomials $P^{(l),k}$. In particular, the coefficients $P_a^{(l),k}(\mu,x)$ are polynomials of degree $x^{k-a}$ in $x$. Since $\mu$ has a gravitational dimension two and $x$ one, their monomials are of the form $x^{k-a-2n}$ where $n$ is an integer. This implies $P_a^{(l),k}(\mu,-x)=(-1)^{k-a} P_a^{(l),k}(\mu,x)$. Focusing again on the non-vanishing LG correlators, we have
\begin{align}
\begin{split}
&u^{J+k-l+1/2}\mathcal{N}_{Jk}^{(l)}\left(R(P_J,(l-k)\bL,s+i/\bL)+R(P_J,(l-k)\bL,s-i/\bL)\right)\\
=&\sum_{a=0}^k{(-1)^{k-a}P_a^{(l),k}(\mu,x)\left(\Op_{Ja}^{(l)}(s+i/\bL)+\Op_{Ja}^{(l)}(s-i/\bL)\right)}\\
=&(-1)^{l-k} 2\cos(2\pi P_J/\bL)\sum_{a=0}^k{P_a^{(l),k}(\mu,x)\Op_{Ja}^{(l)}(s)}
\end{split}
\end{align}
as required.


\subsection{Vanishing correlators and their constraints}\label{sec_van_corr}
We consider the case $l-k>p-J$ for which the LHS of \ref{rel_res} is equal to zero. It is easy to see by recursion over $k$ that all the monomial terms have to vanish independently, resulting in $l+J-p$ constraints
\begin{equation}\label{constraints}
\la\tr\dfrac{Q^{(l),J}(x,M)M^a}{F_l(x,M)^2}\raCL=\Op_{Ja}^{(l)}(x)\qquad\text{for}\quad 0\leq a<l+J-p.
\end{equation}
Solving these constraints allows to determine the polynomials $Q^{(l),J}(x,M)$ of degree $l+J-p-1$ in $M$ that encodes the bulk-boundary resonance transformation. The bulk-boundary MM correlator $\Op_{Ja}^{(l)}$ has been computed in the section \refOld{sec_MM}. It is convenient to introduce the Chebyshev polynomials of the first and second kind,
\begin{equation}
T_k(x)=u^k\cosh k\pi\bL s,\quad U_{k}(x)=u^k\dfrac{\sinh(k+1)\pi\bL s}{\sinh\pi\bL s},
\end{equation}
and to rewrite the expression \ref{expr_ojkl} as
\begin{align}
\begin{split}\label{expr_ojkl_Cheb}
\Op_{Ja}^{(l)}(x)=(-1)^{l-1}\hf\bL^{-2}u^{2J-p-1/2}&\Bigg[\sinh(\pi s/\bL)\ \sum_{\a=-(l-1):2}^{l-1}{\prod_{\superp{\b=-(l-1):2}{\b\neq\a}}^{l-1}(x_\a-x_\b)^{-1}x_\a^a\dfrac{T_{p-J}(x_\a)}{\sinh(\pi\bL s_\a)}}\\
&-u\cosh(\pi s/\bL)\sum_{\a=-(l-1):2}^{l-1}{\prod_{\superp{\b=-(l-1):2}{\b\neq\a}}^{l-1}(x_\a-x_\b)^{-1}x_\a^aU_{p-J-1}(x_\a)}\Bigg]
\end{split}
\end{align}
where we used the expression of $2P_J=1/\bL-(p-J)\bL$ to expand the hyperbolic sine in the numerators. When $a<l+J-p$, the second sum in \ref{expr_ojkl_Cheb} vanishes because of the property \ref{pow_van} and we simply have,
\begin{equation}\label{equ1}
\Op_{Ja}^{(l)}(x)=(-1)^{l-1}\hf\bL^{-2}u^{2J-p-1/2}\sinh(\pi s/\bL)\ \sum_{\a=-(l-1):2}^{l-1}{\prod_{\superp{\b=-(l-1):2}{\b\neq\a}}^{l-1}(x_\a-x_\b)^{-1}\ \dfrac{x_\a^a T_{p-J}(x_\a)}{\sinh(\pi\bL s_\a)}}.
\end{equation}

The general boundary 2pt functions have been studied in \cite{Bourgine2010b}, but a simplification occurs when the two boundaries are identical. It can be seen by looking at the following correlator,
\begin{equation}
\la\tr\dfrac{M^a}{(M-y)F_l(x,M)}\raCL=-\sum_{\a=-(l-1):2}^{l-1}{\prod_{\superp{\b=-(l-1):2}{\b\neq\a}}^{l-1}(x_\a-x_\b)^{-1} x_\a^a\ \dfrac{\o(y)-\o(x_\a)}{y-x_\a}}.
\end{equation}
When $y$ belongs to the set of $\{x_\a\}_\a$, all the terms vanish apart from $x_\a=y$. For this remaining term, both numerator and denominator cancels and we end up with the resolvent derivative at $x_\a$,
\begin{equation}\label{res343}
\la\tr\dfrac{M^a}{(M-x_\a)F_l(x,M)}\raCL=-\bL^{-2}u^{p-1/2}\prod_{\superp{\b=-(l-1):2}{\b\neq\a}}^{l-1}{(x_\a-x_\b)^{-1}} x_\a^a\ \dfrac{\sinh(\pi s_\a/\bL)}{\sinh\pi\bL s_\a}.
\end{equation}
This result can be plugged in the expression of the MM correlators with two boundaries, to get
\begin{equation}\label{equ2}
\la\tr\dfrac{M^{a+b}}{F_l(x,M)^2}\raCL=(-1)^l\bL^{-2}u^{p-1/2}\sinh(\pi s/\bL)\ \sum_{\a=-(l-1):2}^{l-1}{\prod_{\superp{\b=-(l-1):2}{\b\neq\a}}^{l-1}(x_\a-x_\b)^{-2}\dfrac{x_\a^{a+b}}{\sinh\pi\bL s_\a}}
\end{equation}

Comparing \ref{equ1} and \ref{equ2}, we propose an ansatz for the values of the polynomials $Q^{(l),J}$ at the points $M=x_\a$:
\begin{equation}\label{422}
Q^{(l),J}(x,x_\a)=-\hf u^{2(j-p)} \prod_{\superp{\b=-(l-1):2}{\b\neq\a}}^{l-1}{(x_\a-x_\b)}\ \left[T_{p-J}(x_\a)+q^{(l),J}(x_\a)\sinh\pi\bL s_\a\right]
\end{equation}
where $q^{(l),J}(x_\a)$ is a polynomial of degree $p-J-1$ in $x_\a$ whose coefficients depend on $s$. This term arises because of the freedom in the identification to add $x_\a^c\sinh\pi\bL s_\a$ with $c<l-1-a$ to $T_{p-j}(x_\a)$ within the sum over $\a$. Thus, the expression of $Q^{(l),J}(x,x_\a)$ contains $p-J$ free parameters that will be determined a posteriori. Knowing $l$ values of $Q^{(l),J}$, we can use the Lagrange interpolation formula to obtain $Q^{(l),J}(x,M)$ as a polynomial of degree $l-1$,
\begin{equation}\label{expr_Q}
 Q^{(l),J}(x,M)=-\hf u^{2(J-p)} F_l(x,M)\sum_{\a=-(l-1):2}^{l-1}{\dfrac{T_{p-J}(x_\a)+q^{(l),J}(x_\a)\sinh\pi\bL s_\a}{M-x_\a}}.
\end{equation}
But we know from scaling arguments that $Q^{(l),J}(x,M)$ must be of degree $l-1-(p-J)$, so we have to impose the vanishing of the $p-J$ highest degree terms. These are exactly the number of constraints needed to determine $q^{(l),J}$. These constraints can be derived from the study of the asymptotic at $M\to\infty$,
\begin{equation}\label{suff_cond}
\sum_{\a=-(l-1):2}^{l-1}{\left[T_{p-J}(x_\a)+q^{(l),J}(x_\a)\sinh\pi\bL s_\a\right]\ x_\a^n}=0,\quad n=0\cdots p-1-J.
\end{equation}
To solve these constraints, we first note that we can replace $x_\a^n$ in the previous equations by a Chebyshev polynomial of the first kind with the same degree. Then we decompose $q^{(l),J}$ over a basis of Chebyshev polynomials of the second kind,
\begin{equation}
q^{(l),J}(x_\a)=\sum_{k=0}^{p-1-J}{\tilde{q}_k^{(l),J}(s) u^{p-J-k} U_k(x_\a)}
\end{equation}
which allows to simplify the constraints
\begin{equation}
\sum_{\a=-(l-1):2}^{l-1}{\left(\cosh(p-J)\pi\bL s_\a\cosh n\pi\bL s_\a+\sum_k{\tilde{q}_k^{(l),J}\sinh(k+1)\pi\bL s_\a}\cosh n\pi\bL s_\a\right)}=0
\end{equation}
and eventually perform the $\a$-summation using
\begin{equation}\label{sum_alpha}
\sum_{\a=-(l-1):2}^{l-1}{\sinh\g\pi\bL s_\a}=\dfrac{s_{\g l}}{s_\g}\sinh\g\pi\bL s,\quad \sum_{\a=-(l-1):2}^{l-1}{\cosh\g\pi\bL s_\a}=\dfrac{s_{\g l}}{s_\g}\cosh\g\pi\bL s.
\end{equation}
We end up with a linear system over $\tilde{q}_k^{(l),J}$,
\begin{equation}\label{syst_Q}
\sum_{k=0}^{p-1-J}\mathcal{U}_{nk}\tilde{q}_k^{(l),J}=\dfrac{-\mathcal{T}_n}{\sinh\pi\bL s},\quad \mathcal{U}_{nk}=\hf\sum_\pm{\dfrac{s_{(k+1\pm n)l}}{s_{k+1\pm n}}U_{k\pm n}(x)},\quad \mathcal{T}_n=\hf\sum_\pm{\dfrac{s_{(p-J\pm n)l}}{s_{p-J\pm n}}T_{p-J\pm n}(x)}
\end{equation}
that can be solved by inverting the $(p-J)\times(p-J)$ matrix $\mathcal{U}(s)$.\footnote{For a matter of clarity, here the vectors and matrix indices start from zero instead of the usual convention one.} This technique provides a unique expression for the polynomials $Q^{(l),J}(x,M)$ that solves the constraints \ref{constraints}. From this expression, it is not obvious that the coefficients of $Q^{(l),J}(x,M)$ are polynomials in $x$ and $\mu$. This can be shown, provided that the determinant of $\mathcal{U}$ cancels with the numerators of each coefficients. This is indeed what happened in the few cases we checked.

As a crosscheck for the expression \ref{expr_Q} of $Q^{(l),J}$, we consider the insertion of a bulk identity operator, $J=p-1$. In this simple case, there is only one constraint that can be easily solved,
\begin{equation}\label{det_q}
q^{(l),p-1}=-\dfrac{u\cosh\pi\bL s}{\sinh\pi\bL s}.
\end{equation}
But since the cosmological constant $\mu=t_{p-1}$ is not turned off at the critical point $\ast$ and is not resonant, $Q^{(l),p-1}$ is just the derivative of the boundary matrix operator $F_l(x,M)$ with respect to $\mu$ at fixed $x$,
\begin{align}
\begin{split}
Q^{(l),p-1}(x,M)&=\left.\dfrac{\p F_l(x,M)}{\p\mu}\right\vert_\text{$x$ fixed}=\left[\dfrac{\p}{\p\mu}-\dfrac{\cosh\pi\bL s}{2\pi\bL\mu\sinh\pi\bL s}\dfrac{\p}{\p s}\right] F_l(x,M)\\
&=\dfrac{F_l(x,M)}{2\sqrt{\mu}\sinh{\pi\bL s}}\sum_{\a=-(l-1):2}^{l-1}{\dfrac{\sinh i\pi\bL^2\a}{M-x_\a}}.
\end{split}
\end{align}
This result is in agreement with the expression \ref{expr_Q} and \ref{det_q}.

As another example, let us investigate the case $J=p-2$ where the matrix $\mathcal{U}$ is simply $2\times 2$. The inversion gives
\begin{equation}
\tilde{q}_0^{(l),p-2}=-\dfrac{c_lc_1}{x^2-u^2c_1^2}\dfrac1{\sinh\pi\bL s},\quad \tilde{q}_1^{(l),p-2}=-\dfrac{2x^3 -(2+c_2)xu^2}{2(x^2-u^2c_1^2)}\dfrac{u^{-1}}{\sinh\pi\bL s}.
\end{equation}
Even in the case of $l=4$, the expression for $Q^{(4),p-2}$ is rather complicated and we will not give it here. However, we have been able to check that the factor $x^{2}-u^2c_1^2$ appearing in the denominator of the two previous coefficients cancels so that each monomial of $Q^{(4),p-2}(M)$ is indeed a polynomial in $x$ and $\mu$. We believe that the cancelation of $\det\mathcal{U}$ is a general feature such that the polynomial $Q^{(l),p-J}(M)$ determined by this method always has the required form.

\subsection{Non-vanishing correlators}\label{sec_nvan_corr}
We investigate here the relation \ref{rel_res} in the case of non-vanishing LG correlators. As already mentioned above, there is no bulk-boundary resonance involved. The non-resonant case $k=0$ has already been treated as a preliminary check. In this section, we restrict ourselves to the cases $k=1,2,l-2$ and $l-1$. The aim is not to derive a general proof but only to provide convincing arguments for the consistency of the approach presented in section \refOld{sec_res_trans} and the pure boundary resonance expression found in \cite{Bourgine2010b}. Accordingly, we will concentrate on the $\mu$- and $s$- dependent part of the LG and MM correlators which are already non-trivial. In particular, we neglect the coefficient $\mathcal{N}_{Jk}^{(l)}$ that contains the matter and ghost sectors contributions. With a more careful analysis, one could also a priori derive their expression from the MM side but this is beyond the scope of this paper.

The pure boundary resonance transformation, encoded in the polynomials $P^{(l),k}(x,M)$, has been determined in \cite{Bourgine2010b} by solving the orthogonality constraints on the boundary 2pt functions. Unfortunately, their explicit calculation is tedious when $k$ is large since it involves the inversion of a matrix of size $(k+1)\times(k+1)$. This is why in appendix \refOld{appD} we developed an alternative derivation which is useful for $k$ close to $l$.

\paragraph{Case $k=1$:} The resonance polynomial is of degree one and reads \cite{Bourgine2010b},
\begin{equation}\label{expr_p1}
P^{(l),1}(x,M)=M-\dfrac{x}{c_{l-1}}
\end{equation}
To compute the MM correlator involving $P^{(l),1}(x,M)$, we need the expression of $\Op_{J1}^{(l)}$. The later can be obtained using the third equation of \ref{equ_shift_MM} to express it in terms of shifted $\Op_{J0}^{(l)}$,
\begin{equation}
\Op_{J1}^{(l)}(s)=\hf\left(\Op_{J0}^{(l-1)}(s+i\bL)+\Op_{J0}^{(l-1)}(s-i\bL)\right)+xc_{l-1}\Op_{J0}^{(l)}(s).
\end{equation}
The sum of the two shifted MM correlators can identified with LG correlators through \ref{k0}, and simplified using \ref{rec_LG},
\begin{equation}
\Op_{J0}^{(l-1)}(s+i\bL)+\Op_{J0}^{(l-1)}(s-i\bL)=-4\dfrac{c_rs_{l-1}^2}{c_{l-1}}u^{J-l+3/2}\mathcal{N}_{J0}^{(l)}R(P_J,(l-1)\bL,s)+2x\dfrac{s_{l-1}^2}{c_{l-1}}\Op_{J0}^{(l)}(s)
\end{equation}
where we denoted $r=2P_J/\bL$. We deduce
\begin{equation}\label{k1}
\la\tr\dfrac{P^{(l),1}(x,M)}{F_l(x,M)}\tr V_J(M)\raCLc=u^{J-l+3/2}\mathcal{N}_{J1}^{(l)}R(P_J,(l-1)\bL,s),\quad \mathcal{N}_{J1}^{(l)}=-2\dfrac{c_rs_{l-1}^2}{c_{l-1}}\mathcal{N}_{J0}^{(l)}.
\end{equation}

\paragraph{Case $k=2$:}
The expression of $P^{(l),2}$ is known explicitly (see the formula (4.13) of \cite{Bourgine2010b}),
\begin{equation}
P^{(l),2}(x,M)=M^2-2\dfrac{c_1}{c_{l-2}}xM+\dfrac{(s_{l-1}+s_{l-2}c_1)}{s_{2l-3}c_{l-2}}x^2-\dfrac{s_1s_{l-1}^2}{s_{2l-3}}u^2.
\end{equation}
To avoid too complicated expressions, we simply show the proportionality of the MM correlators with $R(P_J,(l-2)\bL,s)$ but the coefficient $\mathcal{N}_{J2}^{(l)}$ could also be determined. We notice that the insertion of the polynomial
\begin{equation}
\bar{P}^{(l),2}(x,M)=\dfrac{F_l(x,M)}{F_{l-2}(x,M)}=M^2-2c_{l-1}xM+x^2-u^2s_{l-1}^2
\end{equation}
within the MM correlator trivially gives an expression proportional to $R(P_J,(l-2)\bL,s)$. Consequently, we only need to show that the difference of the two polynomials, now of degree one in $M$, is also proportional to this LG correlator,
\begin{equation}
\hat{P}^{(l),2}(x,M)=xM-\dfrac{4s_{l-2}x^2+(s_{3l-4}+s_l)u^2}{4s_{2l-3}}\propto P^{(l),2}(x,M)-\bar{P}^{(l),2}(x,M).
\end{equation}
It can be rewritten by introducing the polynomial $P^{(l),1}$ of \ref{expr_p1},
\begin{equation}
\hat{P}^{(l),2}(x,M)=xP^{(l),1}(x,M)+\dfrac{c_{l-2}s_{l-1}}{s_{2l-3}c_{l-1}}(x^2-u^2c_{l-1}^2),
\end{equation}
and inserted into to the MM correlator to give a sum of two LG correlators through \ref{k0} and \ref{k1}:
\begin{equation}\label{parent}
\la\tr\dfrac{\hat{P}^{(l),2}(x,M)}{F_l(x,M)}\tr V_J(M)\raCL=u^{J-l+1/2}\mathcal{N}_{J0}^{(l)}\dfrac{s_{l-1}^2}{c_{l-1}}\left[-2xuc_r R(P_J,(l-1)\bL,s)+\dfrac{c_{l-2}(x^2-u^2c_{l-1}^2)}{s_{2l-3}s_{l-1}}R(P_J,l\bL,s)\right].
\end{equation}
The expression inside the parenthesis can be simplified using the recursion relation \ref{rec_LG_b}, it is indeed proportional to $R(P_J,(l-2)\bL,s)$. This shows the validity of \ref{rel_res} with $k=2$.

\paragraph{Boundary identity operator ($k=l-1$):}
All the bulk-boundary LG correlators containing the identity boundary operator $B_{l-1}$ are non-vanishing, regardless of the bulk operator. The simplest way to determine the polynomial $P^{(l),l-1}$ is to use the fact that the boundary cosmological constant is not turned off at the $\ast$-critical point. Since this coupling is not resonant, $P^{(l),l-1}$ is simply the derivative of $F_l$ with respect to $t_{l-1}^{(l)}=-\frac{s_l}{s_1}x$,
\begin{equation}\label{expr_P_bound_id}
P^{(l),l-1}(x,M)=-\dfrac{s_1}{s_l}\dfrac{\p F_l(x,M)}{\p x}=\dfrac{s_1}{s_l}\dfrac{F_l(x,M)}{\sinh\pi\bL s}\sum_{\a=-(l-1):2}^{l-1}{\dfrac{\sinh\pi\bL s_\a}{M-x_\a}}
\end{equation}
Pluging this result into the expression \ref{expr_ojkl} for the MM correlator and using \ref{sum_alpha} to perform the $\a$-summation, we obtain
\begin{equation}
\la\tr\dfrac{P^{(l),l-1}(x,M)}{F_l(x,M)}\tr V_J(M)\raCLc\equskip=u^{J-1/2}\mathcal{N}_{Jl-1}^{(l)}R(P_J,\bL,s),\quad \mathcal{N}_{Jl-1}^{(l)}=-\hf\dfrac{s_1}{s_l}\bL^{-3}\dfrac{\sinh2\pi P_J/\bL}{\sinh2\pi P_J\bL}\sinh2\pi lP_J\bL.
\end{equation}
The expression of $\mathcal{N}_{Jl-1}^{(l)}$ at $l=2$ is in agreement with $\mathcal{N}_{J1}^{(l)}$ given in \ref{k1}.

\paragraph{Case $k=l-2$:} The polynomial $P^{(l),l-2}$ can be computed using the method given in appendix \refOld{appD}. The matrix $\mathcal{U}$ to be inverted is only of size $2\times 2$, and we find
\begin{equation}
\tilde{p}_0^{(l),l-2}=-\dfrac{2s_{2l}xu}{s_{2}\D(x)\sinh\pi\bL s},\quad\tilde{p}_1^{(l),l-2}=\dfrac{s_lu^2}{s_1\D(x)\sinh\pi\bL s},
\end{equation}
with the determinant
\begin{equation}
\D(x)=\det\mathcal{U}=4\dfrac{s_l^2(c_l^2-c_1^2)}{c_1s_2s_3}\left(x^2-u^2c_1^2\right).
\end{equation}
Using these results, we can derive the MM correlator associated to the polynomial $P^{(l),l-2}(x,M)$,
\begin{equation}
\la\tr\dfrac{P^{(l),l-2}(x,M)}{F_{l}(x,M)}\tr V_J(M)\raCLc=u^{J-3/2}\mathcal{N}_{Jl-2}^{(l)}R(P_J,2\bL,s)
\end{equation}
with the expression \ref{expr_P2} for the LG part and ($r=2P_J/\bL$)
\begin{equation}
\mathcal{N}_{Jl-2}^{(l)}=-\dfrac14\bL^{-3}\sinh(2\pi P_J/\bL)\dfrac{c_1s_3(c_1s_rs_lc_{lr}-s_1c_rc_ls_{lr})}{s_1s_rs_l(c_l^2-c_1^2)(c_r^2-c_1^2)}.
\end{equation}
This coefficient is in agreement with the formula \ref{k0} for $l=2$ and \ref{k1} for $l=3$.

\section{Summary and discussion}
In this paper we investigated the bulk-boundary disc correlators of the hermitian matrix model and the $(~2~,~2~p~+~1~)$ minimal Liouville gravity. In our study of LG, we specialized to the case of a coupling to degenerate matter operators for which the Liouville dressing charges takes only a finite set of values. We derived two different expression, \ref{expr_R} and \ref{sol_bb_LG}, for the cosmological constants dependent part of the correlator. We also provided various useful relations involving either a shift of the boundary parameter (\ref{LG_shift_1bL},\ref{bulk_rec} and \ref{rec_LG}) or a recursion relation on the boundary operator momentum \ref{rec_LG_b}. Given the large range of applications for the Liouville theory, such relations could also be useful in other contexts.

In the second section, we constructed the MM bulk-boundary correlator following the method of \cite{Ishiki2010,Bourgine2010a}. We obtained its expression in the continuum limit \ref{expr_ojkl} and successfully identified it to the LG correlator in the non-resonant case. But our main interest was in the study of the resonance transformation at the bulk-boundary level. For this purpose, we considered this transformation in a very general setting in section three. It led us to distinguish between two different cases. In the first case, the bulk-boundary LG correlator is vanishing due to the fusion rules obeyed by its matter part. Imposing this vanishing condition on the MM correlators, we were able to find the first order of a bulk-boundary resonance \ref{422}. This transformation can be encoded in a matrix polynomial whose coefficients satisfy a the linear system of equations \ref{syst_Q}.

In the second case, the LG correlator is non-vanishing. There, no bulk-boundary resonance arises - except for the identity operator - and the MM and LG correlators were seen to agree in a number of specific cases, provided we take into account the boundary resonance investigated in \cite{Bourgine2010b}. In this process, we derived an alternative method to compute the matrix polynomials encoding the transformation which is particularly efficient for the boundary operators having a small Kac index. 

The explicit agreement between LG and MM correlators in the non-vanishing case has been shown only for some particular cases. Even though these cases are enough to draw the general pattern, a complete proof is still lacking. Such a proof would provide the expression of the factors $\mathcal{N}_{Jk}^{(l)}$ that can be identified, once a proper normalization is fixed, to the matter contribution of the LG correlator, namely to the bulk-boundary disc correlation function of the $(2,2p+1)$ minimal model.

Our ultimate purpose is to conjecture the expression of the resonance transformation including the boundary effects at all order. Such a proposal was made for the bulk case in \cite{Belavin2009} after the study of the 3-pt functions. The boundary disc 3-pt function can be investigated in a similar way. In particular, the LG correlator is known and obeys the property of factorization into ghost, matter and Liouville sectors. The expression of the Liouville correlator found in \cite{Ponsot2002} is rather complicated but should simplify in the case of a coupling to degenerate matter operators.

We defined the MM boundary perturbations to be simple power of the matrix, in contrast with the bulk case where these perturbations are critical potentials related to the underlying KdV hierarchy. There may exist a better definition of the perturbations in the boundary case, possibly related to an underlying integrable hierarchy. This change of basis for the MM perturbation may allow to rewrite the resonance transformation in a simpler form.

The relations we found in section 1.2 for the LG correlators are believed to be consequences of the presence of the ground ring structures and the connection still need to be specified, as we did for the boundary 2-pt function in \cite{Bourgine2010b}. The exact realization of such structures within matrix models is still an open problem. For instance, in the $O(n)$ model the boundary ground ring identities have been related to the continuum limit of MM loop equations in \cite{Bourgine2009a}. At the moment, no general picture including these two examples has emerged.


Finally, we could also apply this boundary construction to other matrix models, such as the $O(n)$ \cite{Kostov1988}, ADE \cite{Kostov1989} or dimers \cite{Staudacher1989} models. These models have an interpretation as statistical models defined on a fluctuating lattice. We hope to develop a statistical interpretation of the matrix boundaries in this context.

\section*{Acknowledgements}
We would like to acknowledge K. Hosomichi for valuable discussion. JEB thanks the warm hospitality of KEK, Yukawa I. and Tokyo U. where this work was completed.
This work is partially supported by the National Research Foundation of Korea (KNRF) grant funded by the Korea government (MEST) 2005-0049409 (B, I \& R) and R01-2008-000-21026-0 (R). The work of G.I. is supported by the Grant-in-Aid for the Global COE Program ``The Next Generation of Physics, Spun from Universality and Emergence'' from the Ministry of Education, Culture, Sports, Science and Technology (MEXT) of Japan.

\appendix

\section{Technical details for the LG side}
\subsection{Fourier transform}\label{appA}
The asymptotic behavior of the LG correlator when $s\to\infty$ can be deduced from a scaling argument\footnote{The LG correlator obeys a scaling relation that allows to write it in the form $R(P_J,k\bL,s)=\mu_B^{2\s_0}F(\mu/\mu_B^2)$. In the limit $\mu\to0$, the correlator must remain finite so that $F(0)$ is a non-zero constant. In the algebraic limit $s\to\infty$, or $\mu_B>>\sqrt{\mu}$, $F$ can be expanded around zero and we deduce the asymptotic behavior \ref{asympt}.}
\begin{equation}\label{asympt}
R(P_J,k\bL,s)\sim e^{2\pi\s_0s},\quad \s_0=2P_J-k\bL\in\mathbb{R}.
\end{equation}
The Fourier transform of $R(P_J,k\bL,s)$ properly exists only when $\s_0<0$, i.e. when $2P_J<\bL k$, and we first consider this case. Then, the Fourier transform can be inverted by choosing any straight contour lying in the strip $\Re\s\in]\s_0,-\s_0[$,
\begin{equation}
R(P_J,k\bL,s)=\int_{c-i\infty}^{c+i\infty}{d\s e^{-2\pi\s s}\tilde{R}(P_J,k\bL,\s)},\quad c\in]\s_0,-\s_0[
\end{equation}
and the Fourier transform $\tilde{R}(P_J,k\bL,\s)$ is analytic in this strip. We now examine the poles of the expression \ref{sin_left} for $\tilde{R}$. Since some of the zeros in the denominator cancel with zeros in the numerator, we are left with two infinite series of poles centered on $\pm P_J$, poles being separated by a distance $\bL$. In addition, the two intervals of length $k\bL$, $]\pm P_J-k\bL/2,\pm P_J+k\bL/2[$ are free of poles,
\begin{equation}\label{poles_location}
\s=\pm P_J+\hf k\bL+n\bL,\quad n\geq 0\text{ or }n\leq -k.
\end{equation}
When $\s_0<0$, the two intervals overlap and the strip $\Re\s\in]-\s_0,\s_0[$ around the origin is indeed free of poles. Closing the contour on the right, the sum of residues leads to the formula \ref{expr_R} for the LG correlator. When $\s_0>0$, the two intervals free of poles does not overlap anymore. However, we can still make sense of $\tilde{R}$ as a Fourier transform by smoothly deforming the contour from the case $\s_0>0$ in the inversion formula. This contour, denoted $\mathcal{C}$, is such that only the poles of \ref{poles_location} with $n\geq0$ are picked up. The residue formula again lead us to the expression \ref{expr_R} for the LG correlator. We can check that this expression has the correct asymptotic \ref{asympt} and satisfies the two reflection properties,
\begin{equation}
R(P_J,k\bL,s)=R(-P_J,k\bL,s),\quad R(P_J,k\bL,s)=R(P_J,k\bL,-s).
\end{equation}

\subsection{Derivation of the shift equation}\label{appB}
To derive the two shift relations \ref{rec_LG}, we first notice that $\d_{\pm}$ is invariant under $k\to k+1$, $\s\to \s+\bL/2$ and since the shift of $k$ increment the number of terms in the numerator product of \ref{sin_left}, we get
\begin{equation}
\tilde{R}(P_J,(k+1)\bL,\s)=4\sin\pi\bL(\s+(k-1)\bL/2+P_J)\sin\pi\bL(\s+(k-1)\bL/2-P_J)\ \tilde{R}(P_J,k\bL,\s-\bL/2)
\end{equation}
This relation can be inserted within the Fourier transform, and the shift over $\s$ can be reabsorbed under a change of variable $\s\to\s-\bL/2$. This change of variable shifts the contour of $\bL/2$ to the left, so that we could a priori pick up an extra pole. However, since the poles of $\tilde{R}$ in $\s$ are separated by a distance $\bL$ and $J$, $k$ are integer, the contour $\mathcal{C}$ can always be chosen such that no additional poles are picked up. This choice will be confirmed below by the reflection properties $P_J\to-P_J$ of the recursion relation, and we can safely write
\begin{equation}
R(P_J,(k+1)\bL,s)=-ie^{-\pi\bL s}\int_\mathcal{C}{d\s e^{-2\pi s\s}\ 4\sin\pi\bL(\s+k\bL/2+P_J)\sin\pi\bL(\s+k\bL/2-P_J)\ \tilde{R}(P_J,k\bL,\s)}.
\end{equation}
Then, we use a trigonometric identity to transform the product of sines into a difference of cosines,
\begin{equation}
R(P_J,(k+1)\bL,s)=ie^{-\pi\bL s}\int_\mathcal{C}{d\s e^{-2\pi s\s}\left[e^{2i\pi\bL \s}e^{i\pi\bL^2k}+e^{-2i\pi\bL \s}e^{-i\pi\bL^2k}-2\cos2\pi\bL P_J \right]\tilde{R}(P_J,k\bL,\s)}.
\end{equation}
Inside the brackets, the first two terms can be absorbed by a shift of the variable $s$, the third one does not depend on the integration variable, so we end up with
\begin{equation}\label{equ_shift_LG}
e^{\pi\bL s}R(P_J,(k+1)\bL,s)=2\cos2\pi\bL P_J\ R(P_J,k\bL,s)-e^{i\pi\bL^2k}R(P_J,k\bL,s-i\bL)-e^{-i\pi\bL^2k}R(P_J,k\bL,s+i\bL)
\end{equation}
Note that this relation is invariant under the bulk reflection $P_J\to -P_J$ which confirms the fact that no extra poles were picked up since individual residues does not obey this symmetry. Finally, using the property $R(P,\b,-s)=R(P,\b,s)$ we easily derive the two equations \ref{rec_LG}.

\subsection{Solution of the second shift equation \ref{rec_LG}}\label{appC}
To show that the expression \ref{sol_bb_LG} satisfies the recursion relation given by the second equation of \ref{rec_LG}, we just plug it in the RHS of this equation and then shift the indices $\a$ to get
\begin{align}
\begin{split}
R(P_J,(k+1)\bL,s)&=u^{k-1}\dfrac{\G_k\sinh i\pi\bL^2k}{\sinh\pi\bL s}\sum_\pm{\pm\sum_{\a=-(k-1):2}^{k-1}{\prod_{\superp{\b=-(k-1):2}{\b\neq\a}}^{k-1}{(x_{\a\pm1}-x_{\b\pm1})^{-1}}\ \dfrac{\sinh2\pi P_Js_{\a\pm1}}{\sinh\pi\bL s_{\a\pm1}}}}\\
&=u^{k-1}\dfrac{\G_k\sinh i\pi\bL^2k}{\sinh\pi\bL s}\Bigg(\sum_{\a=-(k-2):2}^{k}{\prod_{\superp{\b=-(k-2):2}{\b\neq\a}}^{k}{(x_{\a}-x_{\b})^{-1}}\ \dfrac{\sinh2\pi P_Js_{\a}}{\sinh\pi\bL s_{\a}}}\\
&\quad-\sum_{\a=-k:2}^{k-2}{\prod_{\superp{\b=-k:2}{\b\neq\a}}^{k-2}{(x_{\a}-x_{\b})^{-1}}\ \dfrac{\sinh2\pi P_Js_{\a}}{\sinh\pi\bL s_{\a}}}\Bigg)
\end{split}
\end{align}
where we denoted
\begin{equation}
\G_k=-\dfrac{2^{k-1}\bL}{\sinh(2\pi P_J/\bL)}\prod_{\g=1}^{k-1}{\sinh^2 i\pi\bL^2\g}.
\end{equation}
We first examine the term $\a=k$ appearing in the first sum and see that in order to form a complete product over $\b$ running form $-k$ to $k$ with steps of two (excepting $\a$), we need an extra factor $(x_k-x_{-k})$. Similarly the same extra factor is needed for the term $\a=-k$. Then, all the other terms have contributions of both sums which differences leads to the same factor:
\begin{equation}
\prod_{\superp{\b=-(k-2):2}{\b\neq\a}}^{k}{(x_{\a}-x_{\b})^{-1}}-\prod_{\superp{\b=-k:2}{\b\neq\a}}^{k-2}{(x_{\a}-x_{\b})^{-1}}=(x_k-x_{-k})\prod_{\superp{\b=-k:2}{\b\neq\a}}^{k}{(x_{\a}-x_{\b})^{-1}}
\end{equation}
Thus, the correlator with shifted $\b$ is equal to
\begin{align}
\begin{split}
R(P_J,(k+1)\bL,s)&=\dfrac{\G_k\sinh i\pi\bL^2k}{\sinh\pi\bL s} (x_k-x_{-k})\sum_{\a=-k:2}^{k}{\prod_{\superp{\b=-k:2}{\b\neq\a}}^{k}{(x_{\a}-x_{\b})^{-1}}\ \dfrac{\sinh2\pi P_Js_{\a}}{\sinh\pi\bL s_{\a}}}\\
&=u^{k}\G_{k+1} \sum_{\a=-k:2}^{k}{\prod_{\superp{\b=-k:2}{\b\neq\a}}^{k}{(x_{\a}-x_{\b})^{-1}}\ \dfrac{\sinh2\pi P_Js_{\a}}{\sinh\pi\bL s_{\a}}}.
\end{split}
\end{align}

\section{Alternative solution for the pure boundary resonance encoding polynomial}\label{appD}
The solution found in \cite{Bourgine2010b} for $P^{(l),k}$ is manageable when $k$ is small since it involves the computation of determinants of size $(k+1)\times(k+1)$. The alternative solution we demonstrate here is convenient when $l-k$ is small, relying on the inversion of a matrix of size $(l-k)\times(l-k)$. However this solution is specific to the case of two identical boundaries. The orthogonality conditions arising from the boundary 2pt function can be written as \cite{Bourgine2010b},
\begin{equation}\label{B1}
\la\tr\dfrac{P^{(l),k}(x,M)M^a}{F_l(x,M)^2}\raCL=0,\qquad \text{for  }\ 0\leq a<k.
\end{equation}
Together with the constraint that $P^{(l),k}(x,M)$ is a monic polynomial of degree $k$ in $M$, these requirements fully determine $P^{(l),k}$. The MM correlators appearing in \ref{B1} can be computed using equ. \ref{equ2},
\begin{equation}
\sum_{\a=-(l-1):2}^{l-1}{\prod_{\superp{\b=-(l-1):2}{\b\neq\a}}^{l-1}(x_\a-x_\b)^{-2}\dfrac{P^{(l),k}(x,x_\a)x_\a^a}{\sinh\pi\bL s_\a}}=0,\quad \forall 0\leq a<k.
\end{equation}
The following ansatz for the values of $P^{(l),k}(x,x_\a)$ solves the orthogonality constraints,
\begin{equation}
P^{(l),k}(x,x_\a)=\prod_{\superp{\b=-(l-1):2}{\b\neq\a}}^{l-1}{(x_\a-x_\b)}\sinh\pi\bL s_\a\ p^{(l),k}(x_\a),
\end{equation}
where $p^{(l),k}(x_\a)$ is a polynomial of degree $l-1-k$ in $x_\a$ with coefficients depending on $x$ to be determined. Using the Lagrange interpolation formula, we define the polynomials $P^{(l),k}(x,M)$ as
\begin{equation}\label{expr_P}
P^{(l),k}(x,M)=F_l(x,M)\sum_{\a=-(l-1):2}^{l-1}{\dfrac{\sinh\pi\bL s_\a}{M-x_\a}p^{(l),k}(x_\a)}.
\end{equation}
These polynomials are of degree $l-1$ but from scaling arguments $P^{(l),k}$ should be of degree $k$. Thus, we have to impose the vanishing of the $l-k-1$ first coefficients. These $l-k-1$ new constraints are not sufficient to determine the polynomials $p^{(l),k}(x_\a)$ depending on $l-k$ coefficients. The remaining constraint comes from the condition that $P^{(l),k}$ is monic which also restricts its term of degree $k$. All these constraints can be read from the asymptotic behavior at $M\to\infty$,
\begin{equation}\label{req}
\sum_{\a=-(l-1):2}^{l-1}{x_\a^n\ p^{(l),k}(x_\a)\ \sinh\pi\bL s_\a}=\d_{n,l-k-1},\quad 0\leq n<l-k.
\end{equation}
To determine $p^{(l),k}$, we first note that in the previous equation $x_\a^n$ can be replaced by any monic polynomial of the same degree and we choose (suitably normalized) Chebyshev polynomials of the first kind. Then, we decompose $p^{(l),k}$ on a basis of Chebyshev polynomials of the second kind,
\begin{equation}\label{expr_pt}
p^{(l),k}(x_\a)=\sum_{m=0}^{l-k-1}{\tilde{p}_m^{(l),k}\ u^{-(l-k-1+m)} U_{m}(x_\a)}
\end{equation}
in order to rewrite \ref{req} as\footnote{Due to normalization issues for the Chebyshev polynomials, this expression is only valid for $k<l-1$. When $k=l-1$, the RHS is $\d_{n,0}$ instead of $2^{-1}\d_{n,0}$.}
\begin{equation}
\sum_{m=0}^{l-k-1}{\tilde{p}_m^{(l),k}\sum_{\a=-(l-1):2}^{l-1}{\cosh n\pi\bL s_\a\ \sinh(m+1)\pi\bL s_\a}}=2^{l-k-2}\d_{n,l-k-1},\quad 0\leq n<l-k.
\end{equation}
The $\a$-summation can be performed using \ref{sum_alpha} leading to a linear system of equations of $\tilde{p}_m^{(l),k}$ that can be solved by inversion of the matrix $\mathcal{U}$ of size $(l-k)\times(l-k)$:
\begin{equation}
\sum_{m=0}^{l-1-k}{\mathcal{U}_{nm}(x)\tilde{p}_m^{(l),k}(x)}=\dfrac{2^{l-k-2}}{\sinh\pi\bL s}\d_{n,l-k-1},\quad \mathcal{U}_{nm}(x)=\hf\sum_{\pm}{\dfrac{s_{(m+1\pm n)l}}{s_{m+1\pm n}} U_{m\pm n}(x)}.
\end{equation}
This matrix $\mathcal{U}$ is the same that appeared in \ref{syst_Q} for the determination of the bulk-boundary resonance encoding polynomials $Q^{(l),J}$.

As a crosscheck, we can consider the case of the insertion of the boundary identity operator $k=l-1$. The relation \ref{req} supplies only one constraint that can be solved, leading to
\begin{equation}
\tilde{p}^{(l),l-1}=\dfrac{s_1}{s_l}\dfrac{1}{\sinh\pi\bL s}
\end{equation}
which indeed gives a polynomial $P^{(l),l-1}$ that corresponds to the expression \ref{expr_P_bound_id} obtained as a derivative of $F_l$ with respect to $x$. As another crosscheck, the explicit expression for the polynomial with $l=4$ and $k=2$ has also been derived, it matches the formula (4.13) given in \cite{Bourgine2010b}.

\bibliographystyle{elsarticle-num.bst}

\end{document}